\documentclass[submission, Phys]{SciPost}

\begin{document}

\begin{center}{\Large \textbf{
Mapping correlations and coherence: adjacency-based approach to data visualization and regularity discovery}}\end{center}

\begin{center}
Guang-Xing Li\textsuperscript{1},

\end{center}

\begin{center}
{\bf 1} South-Western Institute of Astronomical Research, Kunming, Yunnan,
China, 650500

* ligx.ngc7293@gmail.com
\end{center}

\begin{center}
\today
\end{center}


\section*{Abstract}
{
The development of science has been transforming man's view towards nature for
centuries. Observing structures and patterns in an effective approach to
discover regularities from data is a key step toward theory-building. With
increasingly complex data being obtained, revealing regularities systematically
has become a challenge.
Correlation is a most commonly-used and effective approach to describe
regularities in data, yet for complex patterns, spatial inhomogeneity and
complexity can often undermine the correlations. We present a algorithm
to derive maps representing the type and degree of correlations, by taking the
two-fold symmetry of the correlation vector into full account using the Stokes
parameter. The method allows for a spatially-resolved view of the nature and
strength of correlations between physical quantities. In the correlation view, a
region can often be separated into different subregions with different types of
correlations. Subregions correspond to physical regimes for physical systems, or
climate zones for climate maps.
The simplicity of the method makes it widely applicable to a variety of data,
where the correlation-based approach makes the map particularly useful in
revealing regularities in physical systems and alike. As a new and efficient
approach to represent data, the method should facilitate the development of new computational approaches to regularity discovery.
}

\vspace{10pt}
\noindent\rule{\textwidth}{1pt}
\tableofcontents\thispagestyle{fancy}
\noindent\rule{\textwidth}{1pt}
\vspace{10pt}

linenumbers

\section{Introduction}{}
A great amount of our knowledge of nature comes from making informed analyses of data and
facts. With the development of technology, today's scientific investigations
often involve the analysis of data with an ever-increasing level of
complexity.
Correlation is the simplest, most effective, and most commonly accepted approach to describe the relationship between quantities, and finding correlations is the simplest approach
to revealing regularities from data. In realistic, complex systems,  finding
correlations can become challenging, since both spatial inhomogeneity and
 complexity can undermine correlations. 

One valuable observation, as demonstrated by our related paper, is that for
physical and systems alike, the values measured in a continuous patch in the real space,
should appear continuous in the phase space, and the measured values can often stay correlated. This phenomenon of \emph{locally-correlated variations} is common among a wide range of regular data. However, in most studies, due to a lack of an effective approach to extract local correlations, the information they contain remains mostly unnoticed.

In a related paper (Li 2025 submitted), we proposed to treat these local
correlations as spin-2 vectors and use the Stokes parameters to
superimpose these local correlations. This approach, called the  \emph{adjacent correlation analysis},  has allowed us to construct phase plots, with local and global correlations simultaneously represented. These correlation vectors in the phase space have enabled a dynamical-system-like view of the phase space constructed from image data. In this paper, we present a new method called the \emph{adjacent correlation map}, which presents a spatially resolved view of the correlations at different locations. This allows for the separation of the systems into distinct patches with different properties, as characterized by the way different quantities are correlated. 

\begin{figure}
    \includegraphics[width=\textwidth]{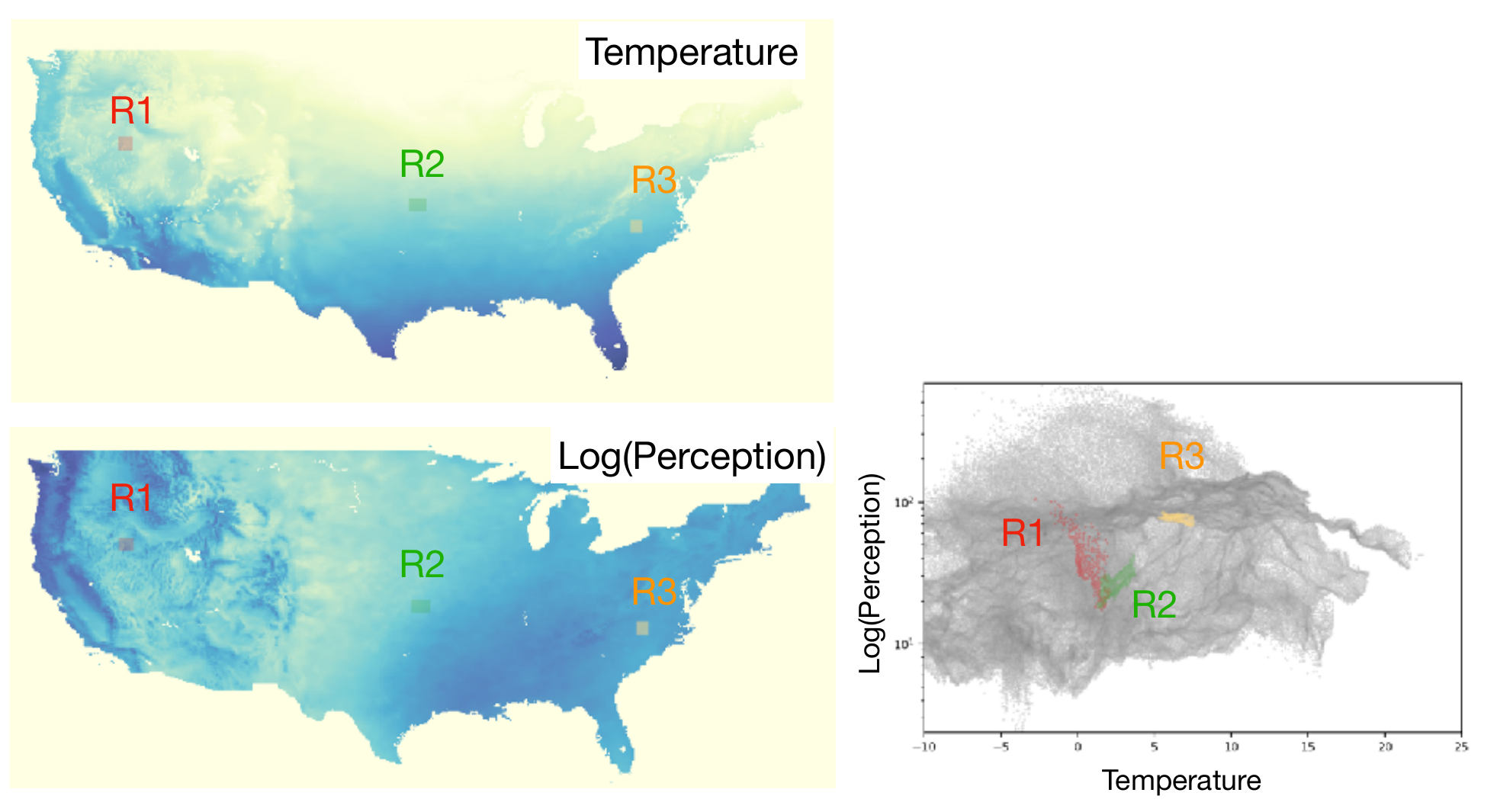}
    \caption{\label{fig:correlation} {\bf Locally-induced correlations }. In the left panels, we plot the temperature ($T$, in Celsius) and the logarithm of perception (${\rm log}(P)$, in mm) measured toward North America. In the right panel, we plot their correlations. \emph{Globally, ${\rm log} P$ and $T$ are not correlated, yet in local regions (R1, R2, and R3), they exhibit negative, positive, and weak correction (dominated by temperature variations). The data is obtained from NOAA, which is measured in January, over the period of 2006 to 2020.
    } 
    \label{fig:local:correlation}}
\end{figure}
 \section{Locally-Correlated Variations: Regularizes from spatially-inhomogeneous systems}
 To illustrate the phenomenon of locally correlated variations, we use measurements of temperature and precipitation from then North America, both of which exhibit complex spatial patterns (Fig. \ref{fig:local:correlation}). When plotted together, we reveal a phase space where temperature and precipitation are not well correlated. To reveal regularities, we choose three boxes ($R1$, $R2$, and $R3$) at different locations. From the west to the east, the temperature and precipitation exhibit correlations ranging from negative, to positive, then to weak correlations. These local correlations are undermined in the global plot and are hard to reveal otherwise.

\begin{figure*}
    \includegraphics[width=1 \textwidth]{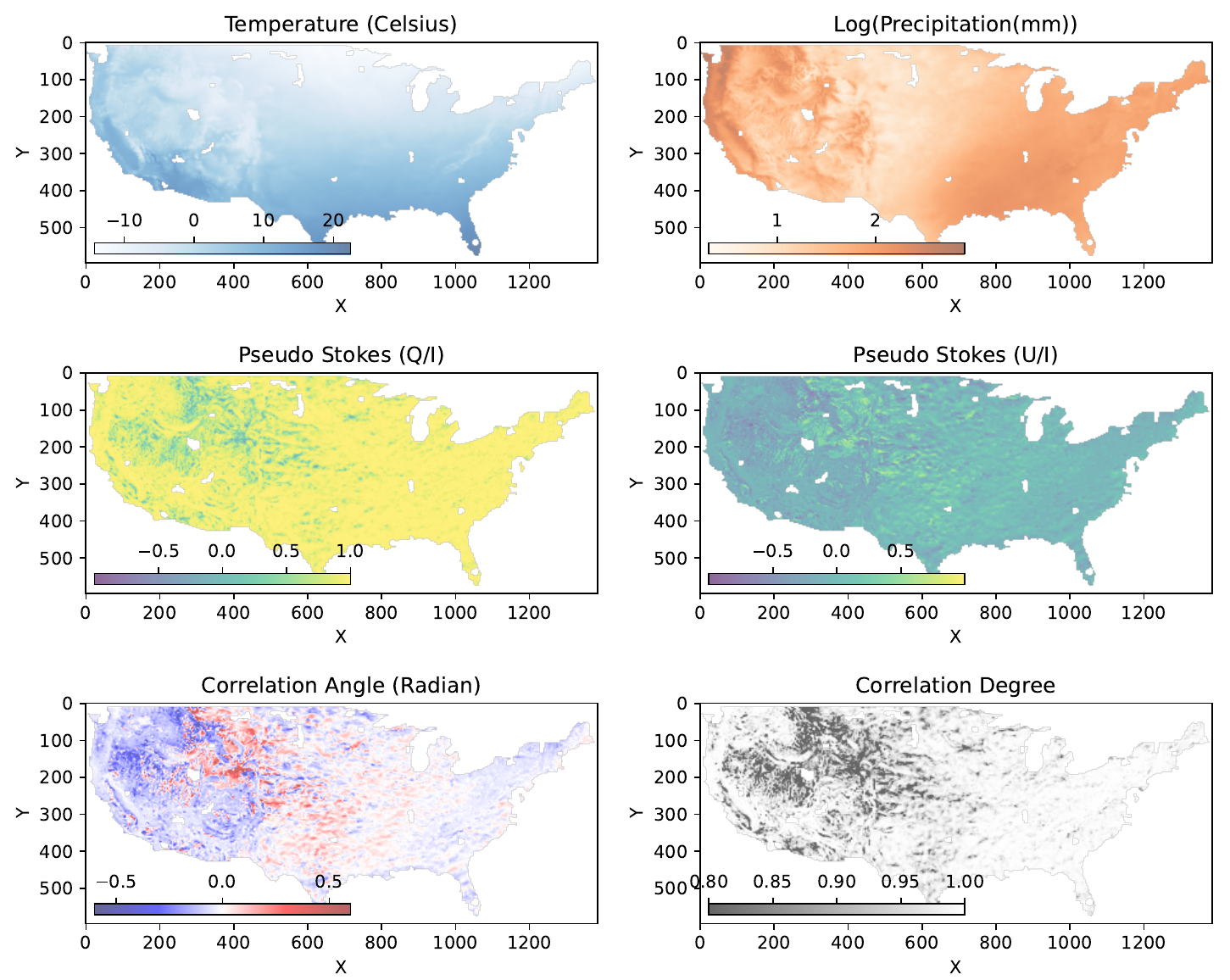}
    \caption{A map of the correlations between the logarithm of the precipitation, measured toward North America, against the temperature. The data was obtained from NOAA. The upper left panel shows the map of the temperature (in Celsius), the upper middle panel shows the map of the logarithm of the precipitation, measured in mm, the middle panels show the map of the Pseudo Stokes parameters, $Q/I$ and $U/I$, and the
 bottom panel shows the map of the correlation angle between temperature $T$ and ${\rm log}(P)$, defined as $\theta = \arctan({\rm d\,log}(P)/ {\rm d} T)$, as well as a map of the correction degree $p$.   \label{fig:working} }
\end{figure*}

\begin{figure*}
    \includegraphics[width=1 \textwidth]{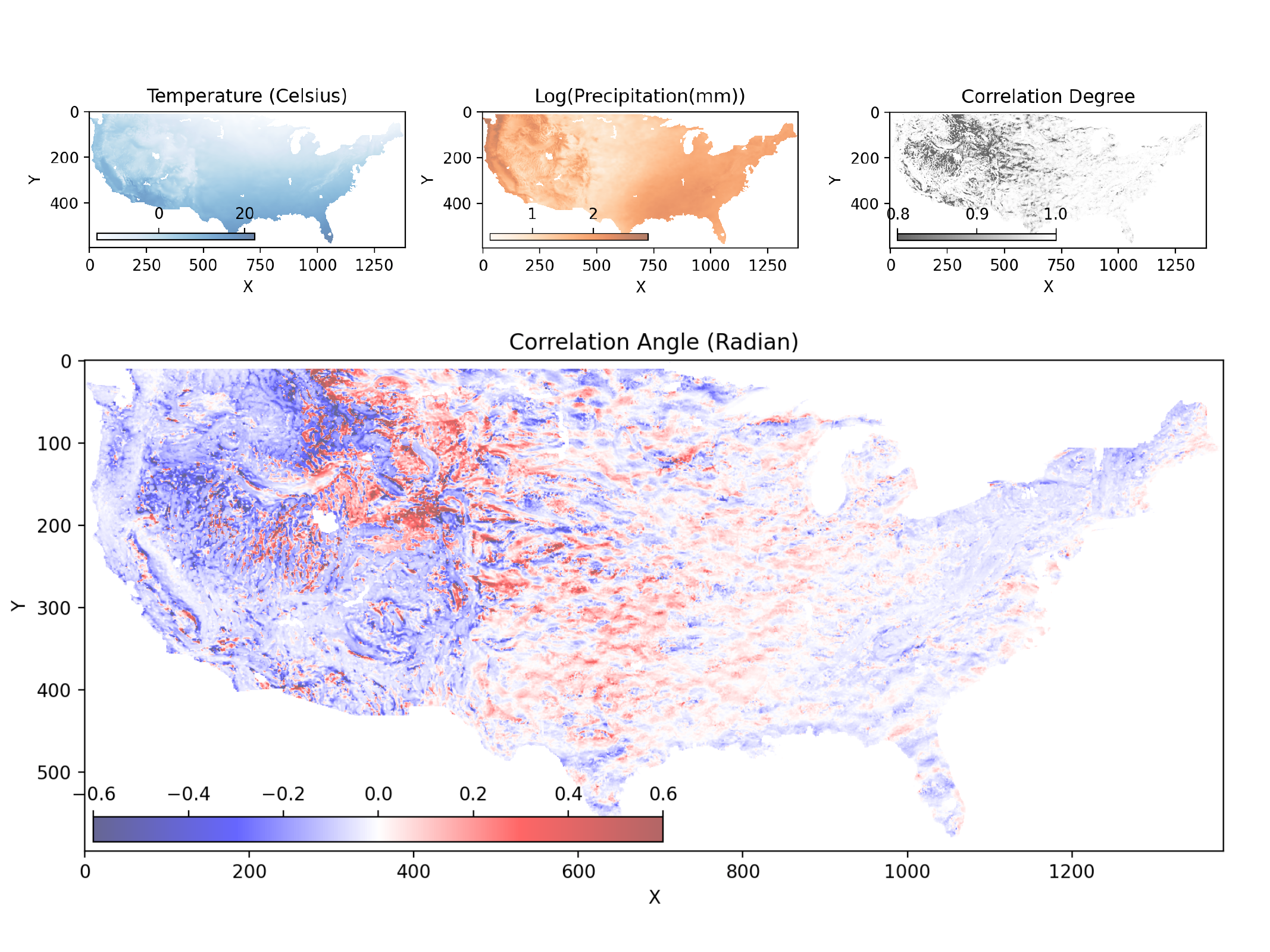}
    \caption{A map of the correlations between the logarithm of the precipitation, measured toward North America, against the temperature. The data are obtained from the NOAA. The upper left panel shows the map of the temperature (in Celsius), the upper middle panel shows the map of the logarithm of the precipitation, measured in mm, the bottom panel shows the map of the correlation angle ($\theta = {\rm arctan} ({{\rm d\,log}}(P)/ {\rm d} T)$) between temperature $T$ and ${\rm log}(P)$. A map of the correction degree is in the upper right corner. Blue color represents regions where ${\rm log}(P)$ correlates negatively with $T$, red regions represent where they correlate positively, and in the white region (in the east) they exhibit weak correlations.  \label{fig:zoom}  }
\end{figure*}

\begin{figure*}
    \includegraphics[width=1 \textwidth]{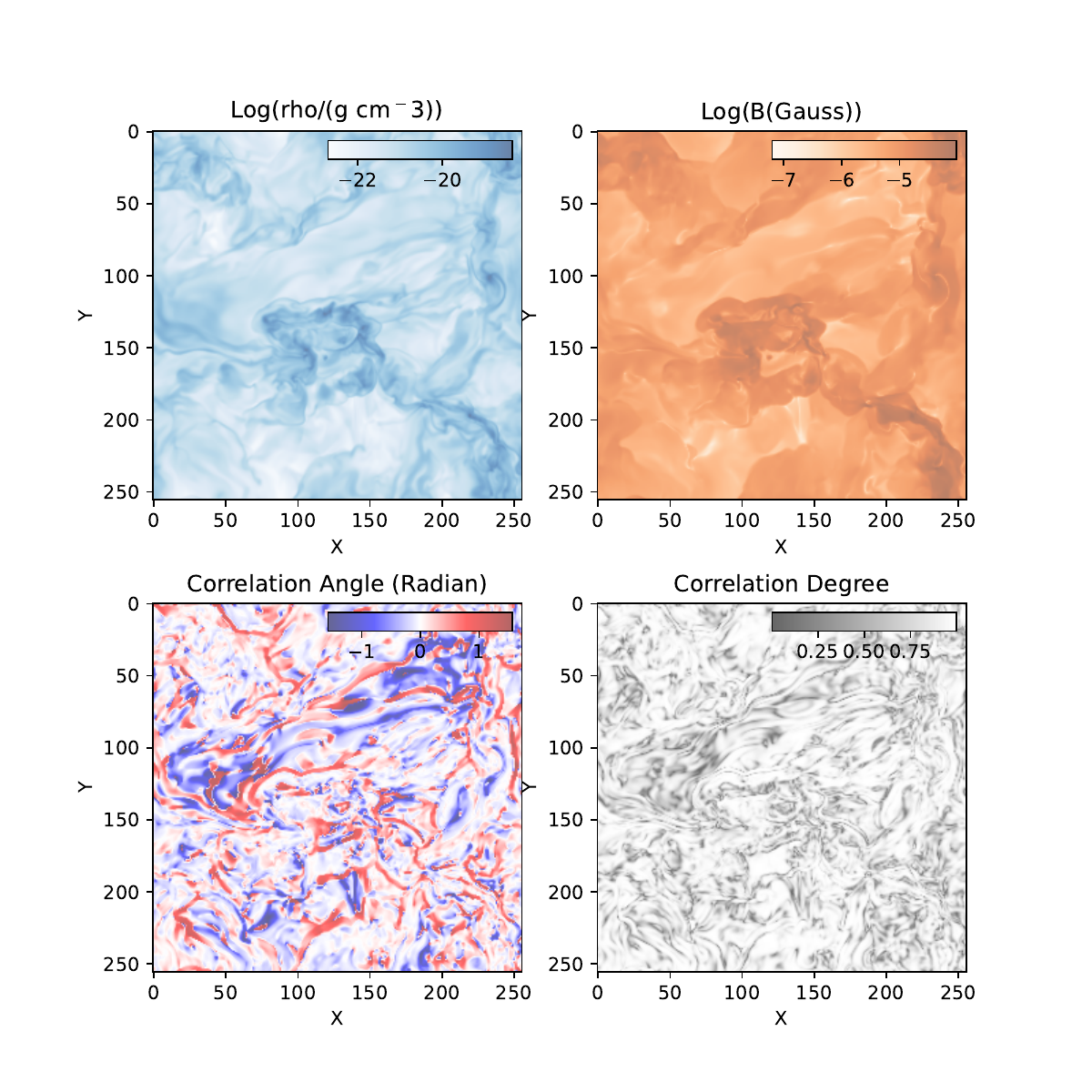}
    \caption{A map of the correlations between the density $\rho$ and the magnetic field $B$ from the simulation of compressive MHD turbulence. The data is obtained from \url{https://www.mhdturbulence.com/}. The upper left panel shows the map of the logarithm of the density (in g/cm$^3$), the upper right panel shows the map of the logarithm of the magnetic field (in Gauss), the bottom panel shows the map of the correlation angle ($\theta = {\rm arctan} ({{\rm d\,log}}(B)/ {\rm d\, log} \rho)$). A map of the correction degree is present in the lower right corner. Blue and red regions are locations where the fluctuations of the logarithm of the magnetic field strength $B$ is much larger than the fluctuations of the logarithm of the density $B$, and the white regions are locations where the fluctuations of the logarithm of the density $\rho$ is much larger than the fluctuations of the logarithm of the magnetic field strength $B$.\label{fig:brho}}
\end{figure*}


%

\begin{figure*}
    \includegraphics[width=1 \textwidth]{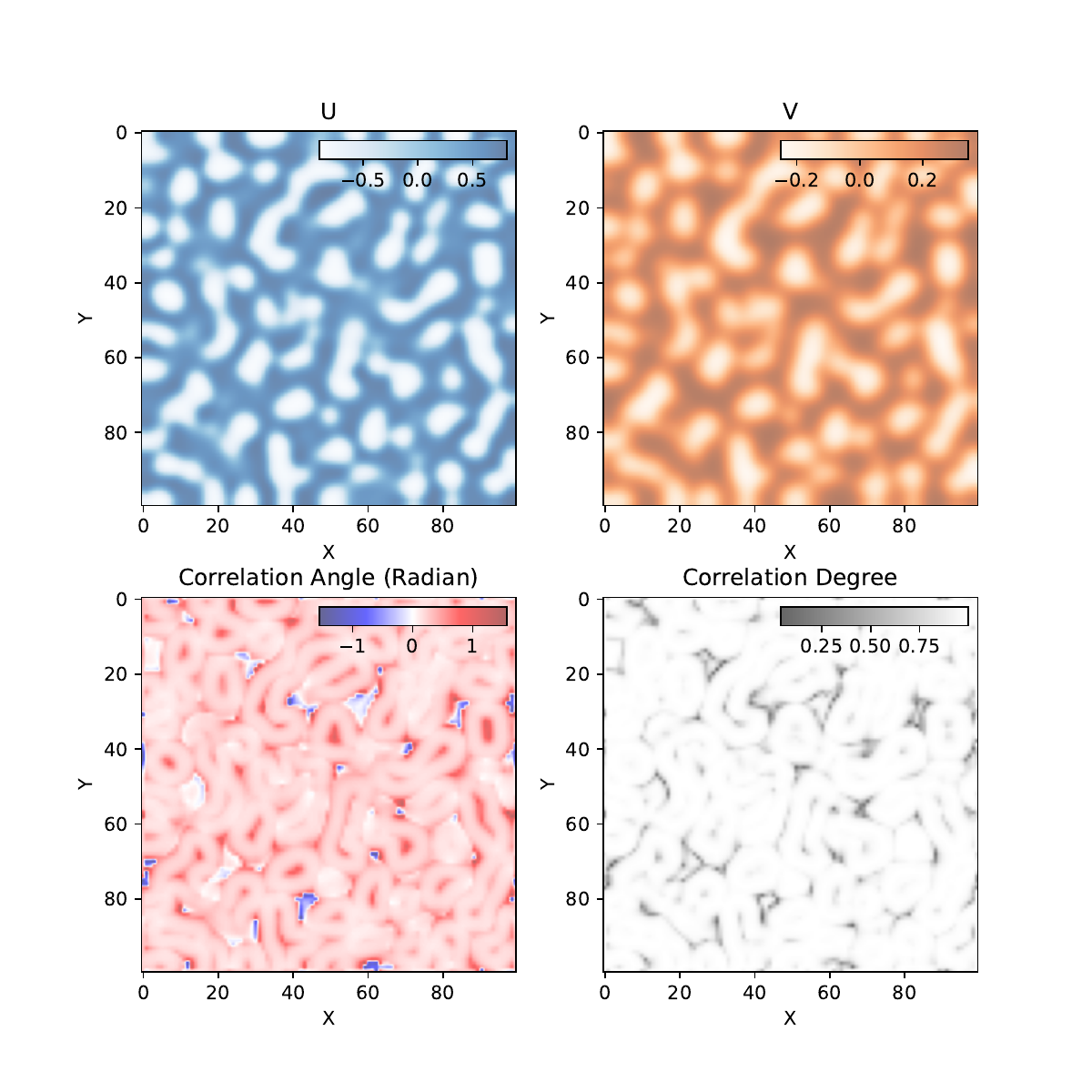}
    \caption{Adjacent Correlation Map of the reaction-diffusion system. In the upper panels, we plot the map of the concentration of the activator, U, and the inhibitor V, and in the lower panels, we plot the map of the correlation angle $\theta = \arctan({\rm d}\, U / {\rm d}\, V)$, and the correction degree. \label{fig:turing} }
\end{figure*}

\section{Adjacent Correlation Analysis}
Although drawing boxes and studying the correlations within can reveal additional regularities, performing such operations can be time-consuming. When the space is occupied by regions of different correlations, it is also difficult to determine the boundary between those.  We further develop a method called the \emph{adjacent correlation map} to reveal such regularities.

\subsection{Mathematical formulation} 
We study the regularities induced by the change of the spatial or temporal coordinates in the space of measurements. 
 We start with a few measurements,
  \begin{equation}
 p_1(\vec{x}), p_2 (\vec{x})\;,
 \end{equation}
 where $p_1, p_2$ are different physical quantities, and  $\vec{x}$
 represents the location.  
At each location, we evaluate the gradients:

\begin{equation}
 \frac{\partial p_{1}}{\partial \vec{x}} = \left( \frac{\partial p_{1}}{\partial x}, \frac{\partial p_{1}}{\partial y}, \frac{\partial p_{1}}{\partial z} \right) \;,     \frac{\partial p_{2}}{\partial \vec{x}} = \left( \frac{\partial p_{2}}{\partial x}, \frac{\partial p_{2}}{\partial y}, \frac{\partial p_{2}}{\partial z} \right) \;,
\end{equation}
and in the phase space, the gradient vectors are

\begin{equation}
{\ {\vec{G_i}}}(\vec{x}) =  (G_{1,i}, G_{2,i}) =\left( \frac{\partial p_{1}}{\partial \vec{x_i}}, \frac{\partial p_{2}}{\partial \vec{x_i}} \right) \;,
   \end{equation}

where the total correlation vector at location $\vec{x}$ is 
\begin{equation}\label{eq:sum}
G(\vec{x}) = \sum_i   {\ {\vec{G_i}}}(\vec{x})\;.
\end{equation}
 
The sum performed in Eq. \ref{eq:sum} needs to be performed assuming that the correlation vectors behave similarly to that of the spin-2 vectors (e.g. they obey the 2-fold rotational symmetry), where we first compute the stokes parameters
\begin{eqnarray}
 I_i(\vec{x}) &=& G_{1,i}(\vec{x})^2 + G_{2,i}(\vec{x})^2\;, \nonumber \\
 Q_i(\vec{x}) &=& G_{1,i}(\vec{x})^2 - G_{2,i}(\vec{x})^2\;, \nonumber \\
 U_i(\vec{x}) &=& 2 G_{1,i}(\vec{x}) G_{2,i}(\vec{x})  \;,
\end{eqnarray}
where
\begin{equation}
 I(\vec{x}) = \sum_i I_i(\vec{x})\;, Q(\vec{x}) = \sum_i Q_i(\vec{x})\;, U(\vec{x}) = \sum_i U_i(\vec{x})\;,
\end{equation}

The stokes $I$ represents the total gradient,  the pseudo-polarization vectors can be computed as 
\begin{equation}
 E_{p_1} = \frac{Q}{I}\;, E_{p_2} = \frac{U}{I}\;,
\end{equation}

\begin{equation}
 p = \left( \left( Q/I\right)^2 + \left(U/I\right)  \right)^{1/2} \;,
\end{equation}
describes how well the two fields, $p_1$ and $p_2$ are correlations, 
and the angle 
\begin{equation}
    \theta = \frac{1}{2} \arctan \left( \frac{U}{Q} \right) \;,
\end{equation}
describes the angle of correlations, e.g. when $p_1$ and $p_2$ are fully correlated, 
\begin{equation}
 {\rm tan}(\theta) = \frac{{\rm d} p_2}{{\rm d} p_1} \;,
\end{equation}
and the correlation degree $p$ describes how well are the two fields correlated: $p=0$ means a non-correction and $p=1$ means a perfect correlation.

To demonstrate the working of the method, we choose data of the temperature and precipitation obtained in North America. In Fig. 
\ref{fig:working}, we present the map of the temperature and the logarithm of the precipitation, maps of the pseudo-stokes $(Q/I)$ and $(U/I)$, as well as the maps of the correlation angle and correlation degree. The both temperature and the precipitation map contain some smooth gradients, which can be seen in the middle of the image, and these gradients, are presented as "regions of a constant correlation" where the color is almost uniform in the map of the pseudo-Stokes parameter $(Q/I)$ and $(U/I)$. This showcases the advantage of using the correlation-based approach, as it can reveal the fact that although the temperature and precipitation values differ from location to location, they can stay correlated in a similar way in a large, coherent region. This extra simplicity is the reason why the adjacent correlation analysis can be an effective approach to data visualization and regularity discovery.

Our Stokes parameter-based formalism shares a direct connection to the correlation matrix, defined as
\begin{equation}
 M_{ij} = <(p_i - \bar{p_i} )(p_j - \bar{p_j})>\;,
\end{equation}
where the use of the Stokes parameters can be viewed as a shortcut to derive the eigenvalues and eigenvectors of the correlation matrix. The correlation degree reflects $l_{\rm max}/({l_{\rm max} + l_{\rm min}})$, where $l_{\rm max}$ and $l_{\rm min}$ are the largest and smallest eigenvalues of the correlation matrix, respectively.  The Stokes I represents the size of the correlation ellipse.
When more than two variables are available, the correlation matrix should be used to derive the correlation vectors.

\section{Applications}
The major advantage of the method is that it is capable of deriving maps representing the correlations between physical quantities. Correlations are often represented in the phase space, which is effective only is the system spatially homogeneous and contamination-free. The adjacent correlation analysis allows for a spatially-resolved view of the corrections between quantities, making is particularly powerful in analyzing data from complex system containing spatially inhomogeneous patches and imperfections. We use some real-world data to demonstrate the effectiveness of the method.



\subsection{Weather data: Precipitation vs. temperature }
We obtain meteorological data from NOAA \footnote{\url{https://www.ncei.noaa.gov/cdo-web/}}. The data are measured in January, over the period of 2006 to 2020, and we apply the \emph{adjacent correlation analysis} to temperature and precipitation data measured toward North America. A zoomed-in view of the correlation angle is plotted in Fig. \ref{fig:zoom}.

From the map, we observe that the temperature and the precipitation appear to be correlated differently at different locations. From the west to the east, they show negative, positive, and weak correlations. A literature survey suggests that this corresponds to these different climate zones, with the western region being semi-arid, and the middle to eastern region being humid continental. In the semi-arid region, water is limited, thus regions with lower temperatures are often associated with high precipitation. In the humid continental region, the water-carrying capacity of the air is the limiting factor. Thus, increases in temperature, which leads to an increase in water-carrying capacity, will lead to an increase in precipitation. This is consistent with the observed positive correlation between temperature and precipitation. In the eastern region, the temperature and precipitation are weakly correlated. These agree broadly with the classification of climate zones in the literature \cite{Trenberth2011-yd}.


\subsection{Simulations of compressive MHD turbulence}
We use numerical simulations of magnetized turbulence
\cite{2012ApJ...750...13C,2015ApJ...808...48B,2020ApJ...905...14B}
\footnote{Available at \url{https://www.mhdturbulence.com/}}. The particular simulation we use has a Mach Number of 6 and a plasma beta=20. As the simulation
runs, the domain is separated into magnetically dominated regions and
kinematically-dominated regions, characterized by different degrees of
magnetization.

We plot the the density ($\rho$) against the magnetic field ($B$), and plot the
result of the adjacent correlation map in Fig. \ref{fig:brho}. The application
of the \emph{adjacent correlation plot} the widespread existence of
locally-correlated variations. Based on the behavior of the system measured in
such correlations, we can further segregate the data into different regimes of
different properties, with density, fluctuate significantly when the magnetic
field dominates, and the magnetic field fluctuates significantly in regions where the
density has been significantly enhanced in the logarithm space. This ratio
between ${\rm d\,log}(B)$ and ${\rm d\, log}(\rho)$ reflects the ratio between
the kinetic energy $E_{\rm k}$ and $E_B$: when the magnetic field dominates, the
density should fluctuate significantly, leading to a correlation angle close to
either $0$ or $\pi$, and when the density dominates, the magnetic field
should fluctuate significantly, leading to a correlation angle close to $\pi/2$ or
$-\pi/2$. The map of the correlation angle and degree allows us to separate the
data into different regions with different properties. Although this behavior
can be seen from the phase plot (Li 2015 submitted), the additional spatial information derived in
the adjacent correlation map allows for a spatially resolved view of the
interplay between density and the magnetic field in detail.

\subsection{Turing pattern from reaction-diffusion systems}
Turing pattern is one of the earliest examples where patterns emerge from a set
of well-defined differential equations and the mechanism of pattern formation
might be responsible for patterns observed in many physical and
biological systems. 

The Turing pattern\cite{1952RSPTB.237...37T} describes how patterns such as stripes and spots can arise naturally and autonomously from a homogeneous, uniform state. It is one of the earliest systems where patterns emerge from a set of well-defined differential equations. 
We simulate the formation of Turing patterns using
\begin{eqnarray} \nonumber
 \frac{\partial u}{\partial t} &=& a \nabla^2 u + u  - u^3 -v + k \nonumber \\
   \tau \frac{\partial v}{\partial t} &=& b  \nabla^2 v + u - v\;, \nonumber 
\end{eqnarray}
where $a=2.8 \times 10^4, b=5\times 10^{-3}, \tau=0.1$ and $k=-0.005$, and study the pattern at the system has reached a quasi-stationary stage.

We apply the \emph{adjacent correlation plot}  and reveal a
different pattern, where the correlation angle at the "shoulder" part of the
activator concentrations appears to be uniform. This can also be seen from the
correlation degree plot, with
a higher correlation degree. We propose this is the region where the activator
and the inhibitor can self-regulate, and acknowledge that more analyses are
needed.


\section{Local emergence of correlations}

Perhaps the most interesting finding from this research is that the
systems often exhibit regularities over large, coherent regions. 

To understand the nature of the correlations, we distinguish between  two type
of regularities: the correlated regularities and the stiff regularities. In
the correlated regularities, two physical quantities will stay correlated
locally, e.g. $\delta(p_1) =  k \delta(p_2)$, and in stiff
regularities, one of the physical quantizes stays constant whereas the other
can vary significantly. 

In Type 1 (correlated) regularity, the physical quantities are correlated,
such that a change of one physical quantity will lead to a change of the other. A
strong Type 1 local correlation implies an emergent simplicity where $p_1 =
f(p_2)$, where different factors can regulate each other. 

In the Type 2 (stiff) regularity, the change of one physical quantity will not lead to a
change of the other. Instead, the local change of one of them dominates over the other. 
 The different degrees of local variations can be related to the stiffness of the governing equations: when the governing equations contain drastically different spatial or temporal scales, the change of fast-changing physical quantity is detached from the slow-changing physical quantity.

Among our examples, 
Type 1 regularities are observed in reaction-diffusion systems, in the western and middle part of North America in the climate data,  where different factors, such as temperature and perception, activator and inhibitor, can self-regulate, and Type 2 regularities are seen in the example of MHD turbulence, where the dominance of the magnetic energy over the kinetic energy leads to a decoupling of the density and the magnetic field, and visa versa.

\subsection{Relation to the Buckinghum's Pi theorem}
Buckingham's Pi Theorem a crucial observation on the relationship between
different physicals. It states that a meaningful physical equation can be written as a function which contains a number of dimensionless groups:
\begin{equation}
f(\Pi_1, \Pi_2, \dots, \Pi_p) = 0
\end{equation}
where $\Pi$ are dimensionless groups, and
\begin{equation}
\Pi_1 = x_1^{a_1} x_2^{a_2} \cdots x_n^{a_n}\;,
\end{equation}
where $x_i$ are the variables, and $a_i$ are the exponents. Since in the  logarithm space, we have
\begin{equation}
 \log(\Pi_1) = \alpha_1 \log(x_1) + \alpha_2 \log(x_2) \;,
\end{equation}
correlations in the log-log space can often be related to the behavior of the system encoded in Buckingham's Pi theorem.

In the cases where the system is simplified, such that 
\begin{equation}
 f(\Pi_1) = 0\;, \Pi_1 = x_1^{a_1} x_2^{a_2} \;,
\end{equation}
we expect 
\begin{equation}
\alpha_1 {\rm d }\,{\rm log } (x_1) + \alpha_2 {\rm d}\,{\rm log} (x_2) = 0\;,  {\rm d }\,{\rm log } (x_1)  =  - \frac{\alpha_2}{\alpha_1} {\rm d}\,{\rm log} (x_2)\;,
\end{equation}
which implies that the two physical quantities are correlated, leading to the \emph{type 1 regularity}.

When \emph{ type 2 regularity} is observed, the local variation of one of the physical quantizes dominates over that of the other. One likely condition is that the quantities belong to different dimensionless groups.

\section{Conclusions}

Discovering regularities from data is a major challenge in research. Correlations are often used to identify regularities. We observe a phenomenon called locally-correlated variations, where physical quantities measured in local regions exhibit stronger correlations than those measured in the entire region. These strong local correlations reflect local regularities, which can be easily undermined by spatial inhomogeneity and noise in the global phase plot. The adjacent correlation map is a systematic approach to reveal such locally-correlated variations while retaining spatial information.

Using real-world data, we demonstrate the widespread existence of locally-correlated variations. We further distinguish between two types of regularities: correlated (Type 1) regularities and stiff (Type 2) regularities. The former are often observed in reaction-diffusion systems, where different factors can self-regulate, and the latter are often observed in MHD turbulence, where density and magnetic fields can vary separately at different locations in the parameter space. Type 1 regularities often imply a form of mutual regulation, while Type 2 regularities often indicate a form of dynamical detachment, where the change in a fast-changing physical quantity is no longer correlated with the behavior of a slow-changing physical quantity due to scale separation.

The method is particularly useful for interactive data exploration, and the new representation it entails facilitates systematic approaches, which will be demonstrated in an accompanying paper.

\bibliography{paper}

\section*{Acknowledgements}
GXL acknowledges support from,
NSFC grant No. 12273032 and 12033005.

\section*{Code Availability}
The code of the method can be found at \url{https://github.com/gxli/Adjacent-Correlation-Analysis}.

\end{document}